\documentclass[iop]{emulateapj}
\usepackage{multirow}
\usepackage{bm}
\usepackage{amssymb}
\usepackage{amsmath}
\usepackage{latexsym}
\usepackage{graphicx}
\usepackage{ifsym}
\usepackage{bm}
\usepackage{tabularx}
\usepackage{hyperref}
\usepackage{enumitem}

\slugcomment{}
\shorttitle{MagAO H$\alpha$ Imaging of HH 508 and LV 1}
\shortauthors{Wu et al.}

\begin{document}
\title{{\textbf {\large T\lowercase{he} I\lowercase{ntricate} S\lowercase{tructure} \lowercase{of} HH 508, \lowercase{the} B\lowercase{rightest} M\lowercase{icrojet} \lowercase{in the} O\lowercase{rion} N\lowercase{ebula}}}}

\author{Ya-Lin Wu, 
 Laird M. Close,
 Jinyoung Serena Kim,
 Jared R. Males, and
 Katie M. Morzinski
}
\affil{Steward Observatory, University of Arizona, Tucson, AZ 85721, USA \vspace{4pt} \\
{\it Published in ApJ}}

\begin{abstract} 
We present {\it Magellan} adaptive optics H$\alpha$ imaging of HH 508, which has the highest surface brightness among protostellar jets in the Orion Nebula. We find that HH 508 actually has a shorter component to the west, and a longer and knotty component to the east. The east component has a kink at 0\farcs3 from the jet-driving star $\theta^1$ Ori $\rm{B}_2$, so it may have been deflected by the wind/radiation from the nearby $\theta^1$ Ori $\rm{B}_1\rm{B}_5$. The origin of both components is unclear, but if each of them is a separate jet, then $\theta^1$ Ori $\rm{B}_2$ may be a tight binary. Alternatively, HH 508 may be a slow-moving outflow, and each component represents an illuminated cavity wall. The ionization front surrounding $\theta^1$ Ori $\rm{B}_2\rm{B}_3$ does not directly face $\theta^1$ Ori $\rm{B}_1\rm{B}_5$, suggesting that the EUV radiation from $\theta^1$ Ori $\rm{C}$ plays a dominant role in affecting the morphology of proplyds even in the vicinity of $\theta^1$ Ori $\rm{B}_1\rm{B}_5$. Finally, we report an H$\alpha$ blob that might be ejected by the binary proplyd LV 1.
\end{abstract}


\section*{\textbf {\normalsize1. I\lowercase{ntroduction}}}
At a distance of $\sim$388 pc, the Orion Nebula is the nearest high-mass star-forming region \citep{K17} and offers a clear view of jet irradiation and disk destruction due to strong UV radiation from OB stars (e.g., \citealt{J98,BR01}), especially in regions around the Trapezium cluster. Observations have revealed the presence of photoevaporated protoplanetary disks (proplyds) with bow-shaped ionization fronts (IFs) facing toward and cometary tails pointing away from the dominant ionizing star of the Trapezium cluster, $\theta^1$ Ori C \citep{O93,OW94,B98,B00}. Over the past $\sim$20 yr, hundreds of proplyds have been imaged from optical to radio wavelengths in the Orion Nebula (e.g., \citealt{B98,Smith05,R08,MW10,Mann14,E16,S16}).

Proplyds are often associated with jets, many of which are short, one-sided, and bright in optical emission, such as H$\alpha$, [\ion{S}{2}], and [\ion{O}{3}]. These compact irradiated jets are sometimes called ``microjets'' \citep{B00,B06,BR01}. \cite{BR01} suggested that the one-sided asymmetry often observed in irradiated jets likely arises from different densities of matter between the irradiated and shadowed sides of a disk. External UV radiation can more effectively remove material on the irradiated side; as a result, the jet facing the UV source can move faster than the counterjet on the shadowed side of the disk. Since the surface brightness of irradiated jets correlates with the amount of incident UV flux when the jets are optically thick to Lyman continuum radiation, jets close to an ionizing star will be more luminous. At $\sim$1\arcsec~from the eclipsing binary $\theta^1$ Ori $\rm{B}_1\rm{B}_5$ (\citealt{Windemuth13}, and references therein), HH 508 is considered the brightest microjet in the Orion Nebula \citep{B00,B06}. In {\it Hubble Space Telescope} ({\it HST}) images, HH 508 appears one-sided and $\sim$$0\farcs5$ in length emerging from $\theta^1$ Ori $\rm{B}_2$. 

We note that $\theta^1$ Ori B (BM Ori) is a hierarchical ``mini-cluster'' with three outer members, $\rm{B}_2\rm{B}_3\rm{B}_4$, orbiting the inner binary $\rm{B}_1\rm{B}_5$ \citep{P98,C03,C12,C13}. \cite{C13} further demonstrated that the components $\rm{B}_2$ and $\rm{B}_3$ also orbit each other in a nearly circular orbit with a period of $\sim$300 yr. The stellar masses for components 1, 2, 3, 4, and 5 are estimated to be 6.3, 3, 2.5, 0.2, and 2.5 $M_\sun$, respectively \citep{V06,C13}. Numerical simulations have shown that this quintuple\footnotemark[1] is unstable, with a very short dynamical lifetime, $\sim$30 kyr \citep{A15,A17}.

\footnotetext[1]{The system might be a sextuple, as \cite{VK04} proposed a third component in an eccentric orbit to the inner binary $\rm{B}_1\rm{B}_5$ to explain the observed radial velocity anomaly.}

This close separation between HH 508 and $\theta^1$ Ori $\rm{B}_1\rm{B}_5$ enables excellent adaptive optics (AO) performance in that the unresolved binary can serve as the natural guide star. It is therefore possible to resolve the structure of HH 508 with ground-based telescopes. In this paper, we present H$\alpha$ images of HH 508 obtained with the {\it Magellan} adaptive optics (MagAO; \citealt{C12,Males14,M14}). MagAO has been fully operational on the 6.5 m Clay Telescope since 2012. With a resolution of 33 mas, we show that HH 508 has complex structure and may have been deflected by the wind and radiation from $\theta^1$ Ori $\rm{B}_1\rm{B}_5$. In addition to HH 508, we also present multi-epoch H$\alpha$ images of the binary proplyd LV 1, showing that the major proplyd may have had episodic mass ejection.

\begin{figure*}[t]
\centering
\includegraphics[angle=0,width=\linewidth]{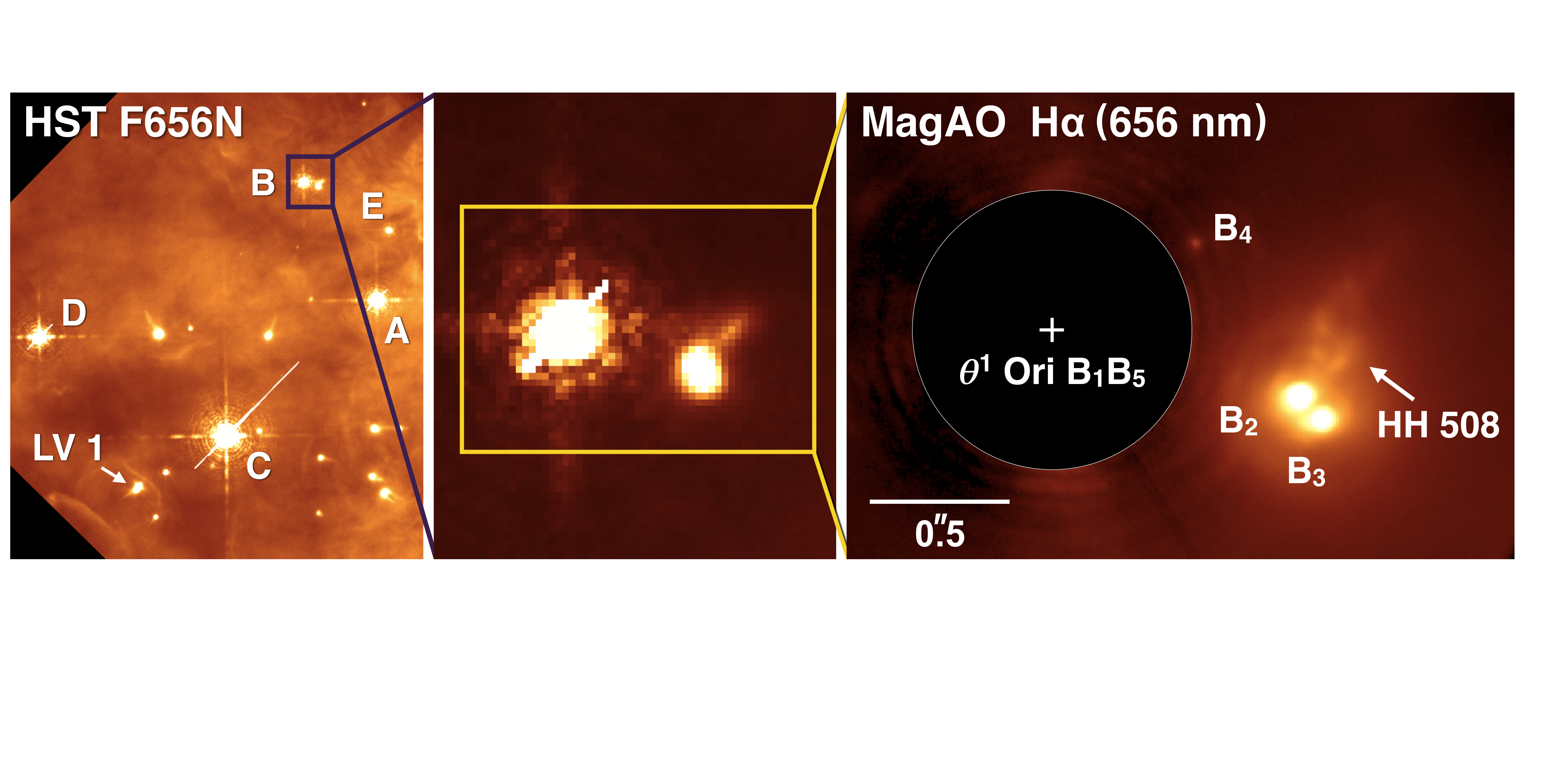}
\caption{MagAO and {\it HST} H$\alpha$ images of HH 508 and the $\theta^1$ Ori B system. We spatially resolve the complex structure of HH 508. The central 0\farcs5~around $\rm{B}_1\rm{B}_5$ in the MagAO image is masked out. The {\it HST} image is from the program GO 5469. North is up and east is left.}
\label{fig:HSTvsMagAO}
\end{figure*}

\begin{deluxetable*}{@{}llrccccc@{}}
\tablewidth{\linewidth}
\tablecaption{MagAO H$\alpha$ Observations of HH 508 and LV 1\label{tab:MagAO_obs}}
\tablehead{
\colhead{Object} &
\colhead{Date} &
\colhead{Speed} &
\colhead{Mode} &
\colhead{Exposure} &
\colhead{Guide Star} &
\colhead{FWHM}
}
\startdata
HH 508				&	2016 Feb 17 	&	990 Hz	&	300	&   2.273 s $\times$ 201	&	$\theta^1$ Ori $\rm{B}_1\rm{B}_5$	&	33 mas\\	\\
LV 1					& 	2012 Dec 03	&	990 Hz	&	250	&   10 s $\times$ 60		&	$\theta^1$ Ori C				&	40 mas\\
					&  	2014 Nov 17	& 	990 Hz	&	300	&   7.5 s $\times$ 488	&	$\theta^1$ Ori C				&	39 mas\\
					&  	2016 Jul 01	& 	990 Hz	&	300	&   60 s $\times$ 66		&	$\theta^1$ Ori C				&	49 mas
\enddata
\end{deluxetable*}

\begin{deluxetable*}{@{}lcccccc@{}}
\tablewidth{\linewidth}
\tablecaption{Luminosities of the Trapezium OB Stars and the UV Fluxes at the Position of HH 508\label{tab:UV}}
\tablehead{
\colhead{Star} &
\colhead{Angular Offset to $\rm{B}_2\rm{B}_3$} &
\colhead{log $L_{\rm bol}$} &
\colhead{log $L_{\rm EUV}$} &
\colhead{EUV Flux at $\rm{B}_2\rm{B}_3$} &
\colhead{log $L_{\rm FUV}$} &
\colhead{FUV Flux at $\rm{B}_2\rm{B}_3$} \\
\colhead{} &
\colhead{(\arcsec)} &
\colhead{($L_\sun$)} &
\colhead{($L_\sun$)} &
\colhead{(W $\rm{m}^{-2}$)}  &
\colhead{($L_\sun$)} &
\colhead{(W $\rm{m}^{-2}$)}}  
\startdata
$\theta^1$ Ori A						&  8.1	&	4.50		&  3.01		&	0.14				&  4.36	&	3.2		\\
$\theta^1$ Ori $\rm{B}_1\rm{B}_5$			&  1.0	&	3.28		&  $-$0.42 	&	$3.5\times10^{-3}$	&  3.05	&	10.5		\\
$\theta^1$ Ori C						&  16.9	&	5.40		&  4.81		&	2.1				&  5.16	&	4.6		\\
$\theta^1$ Ori D						&  20.1	&	4.35		&  2.71		& 	$1.2\times10^{-2}$	&  4.21	&	0.37		\\			
$\theta^1$ Ori E						&  5.3	&	2.42		&  $-$2.85	 	& 	$4.7\times10^{-7}$	&  2.07	&	$3.9\times10^{-2}$		
\enddata
\end{deluxetable*}

\section*{\textbf {\normalsize2. O\lowercase{bservations and} D\lowercase{ata} R\lowercase{eduction}}}
HH 508 was observed in 2016 with $\theta^1$ Ori $\rm{B}_1\rm{B}_5$ ($V$ $\sim$ 6.4 mag) being the guide star, and LV 1 was observed in 2012, 2014, and 2016 with $\theta^1$ Ori C ($V$ $\sim$ 5.1 mag) being the guide star. All data were taken with MagAO's simultaneous differential imaging mode \citep{C14} at 643 nm and 656 nm (H$\alpha$). The 2012 data of LV 1 were previously published in \cite{W13}. Since HH 508 and LV 1 are absent in the 643 nm continuum, we only present the H$\alpha$ images in this paper. Table \ref{tab:MagAO_obs} summarizes our observations. The pixel scale at H$\alpha$ is $\sim$7.9 mas pix$^{-1}$ \citep{C13}. 

The data reduction was carried out with IRAF\footnotemark[2]\footnotetext[2]{IRAF is distributed by the National Optical Astronomy Observatories, which are operated by the Association of Universities for Research in Astronomy, Inc., under cooperative agreement with the National Science Foundation.}(\citealt{T86}, \citeyear{T93}) and MATLAB. We subtracted the dark frame from the raw data, rotated the dark-subtracted frames so that north is up, then registered and combined the images. We removed the halo of $\theta^1$ Ori $\rm{B}_1\rm{B}_5$ from the combined image by subtracting its azimuthal profile.

\begin{figure}[t]
\centering
\includegraphics[angle=0,width=\linewidth]{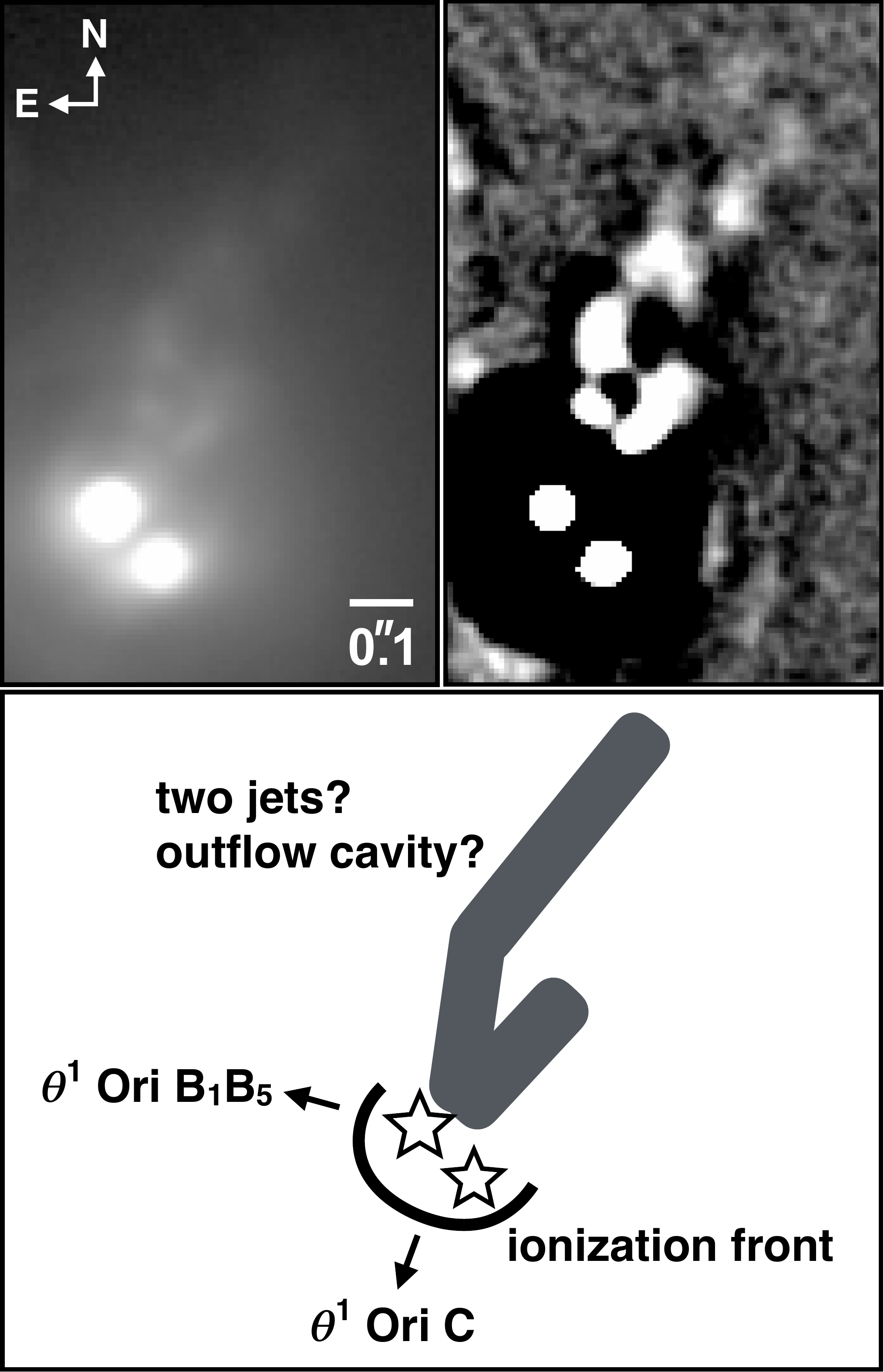}
\caption{We bring out HH 508 by filtering out the halo of the stars with a Gaussian of FWHM = 10 pixels, and further smooth the high-pass filtered image with a two-pixel Gaussian. The east component of HH 508 has many knots and might have been bent by the wind and radiation from $\rm{B}_1\rm{B}_5$. The ionization front surrounding $\rm{B}_2\rm{B}_3$ does not directly face the nearby $\rm{B}_1\rm{B}_5$, indicating that the EUV radiation from $\theta^1$ Ori C dominates the vicinity of $\rm{B}_1\rm{B}_5$.}
\label{fig:HH508_cartoon}
\end{figure}

\begin{figure}[t]
\centering
\includegraphics[angle=0,width=\linewidth]{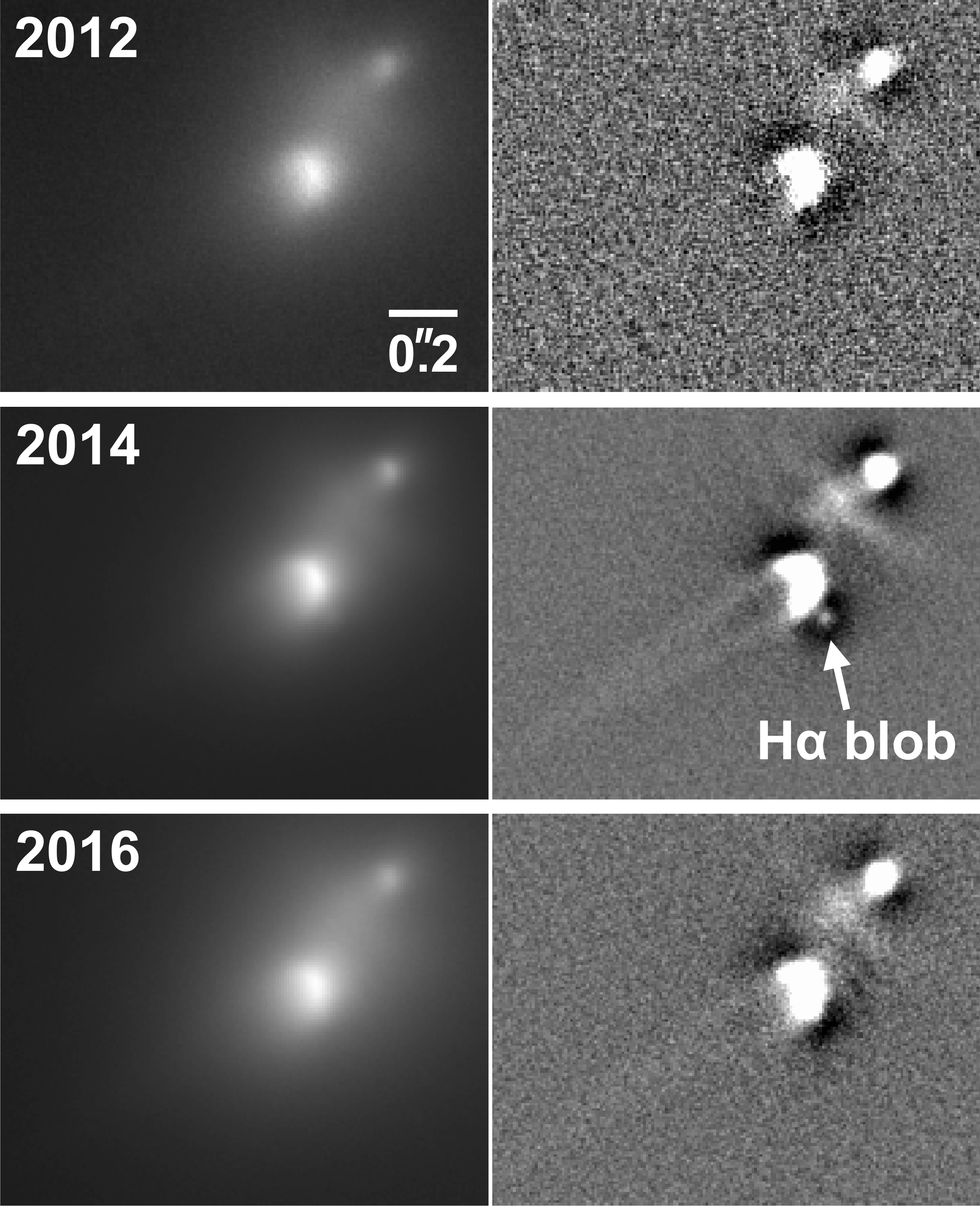}
\caption{H$\alpha$ images of the binary proplyd LV 1 in different epochs. High-pass filtered images are displayed in the right panels. An H$\alpha$ blob emerged to the south of the major proplyd in 2014, but was not seen in the 2016 follow-up imaging.}
\label{fig:LV1}
\end{figure}

\section*{\textbf {\normalsize3. R\lowercase{esults}}}
\subsection*{\textit 3.1. T\lowercase{wo} C\lowercase{omponents of} HH 508}
Figure \ref{fig:HSTvsMagAO} shows the MagAO and {\it HST} H$\alpha$ images of the $\theta^1$ Ori B system. Our high-resolution image reveals a two-component structure for HH 508. The west component is short, while the east component is longer, knotty, and has a kink at $\sim$0\farcs3~from $\theta^1$ Ori $\rm{B}_2$. The kink may imply that the east component has been deflected by $\sim$20\degr~by the wind and radiation from $\theta^1$ Ori $\rm{B}_1\rm{B}_5$. We note that we do not see a cometary tail pointing away from $\rm{B}_1\rm{B}_5$ as mentioned by \cite{B00}. 

The origin of the two-component structure remains puzzling. One possibility is that $\theta^1$ Ori $\rm{B}_2$ is a tight binary, so each component is a separate jet. Alternatively, HH 508 could be a wide and slow-moving outflow, so one or both components may be the outflow cavity walls. Unlike jets, which are collimated and launched in the innermost disk, outflows can be launched at larger disk radii or entrained by the jets. H$\alpha$ outflows have been observed in the Carina Nebula, including HH 666, HH 900, and HH 1163 (\citeauthor{R15a} \citeyear{R15a}, \citeyear{R15b}, \citeyear{R16}). Future spectroscopic or astrometric observations should be able to distinguish fast-moving jets from slow-moving cavity walls.

\subsection*{\textit 3.2. O\lowercase{rientation and} S\lowercase{ize of the} I\lowercase{onization} F\lowercase{ront}}
As shown in Figures \ref{fig:HSTvsMagAO} and \ref{fig:HH508_cartoon}, the bow-shaped ionization front surrounding $\theta^1$ Ori $\rm{B}_2\rm{B}_3$ does not face toward the nearby $\theta^1$ Ori $\rm{B}_1\rm{B}_5$ ($\sim$1\arcsec); in contrast, it points to the distant $\theta^1$ Ori C ($\sim$17\arcsec). Ionizing photons from $\theta^1$ Ori C still affect the morphology of proplyds even in the vicinity of $\theta^1$ Ori $\rm{B}_1\rm{B}_5$. 

The dominant role of $\theta^1$ Ori C can be shown quantitatively by comparing the extreme UV (EUV; $\lambda<912$~\AA) fluxes from the Trapezium OB stars at the position of HH 508. Assuming all of the stars are coplanar and adopting stellar parameters from \cite{G05} and \cite{S05}, we use the model spectra in \cite{CK04} to estimate UV luminosities and fluxes (Table \ref{tab:UV}). We find that the EUV flux ratios are $\rm{A}: \rm{B}_1\rm{B}_5:\rm{C}:\rm{D}:\rm{E} \approx 40:1:600:3:0.0001$. $\theta^1$ Ori C indeed dominates the EUV flux at the position of HH 508 and is responsible for illuminating the IF of $\rm{B}_2\rm{B}_3$.

The IF surrounding $\rm{B}_2\rm{B}_3$ has a radius $R_{\rm IF}\sim60$ au, which is smaller than the average IF radius of $\sim$150 au for the 135 Orion proplyds reported in \cite{VA05}. \cite{K16} showed that $R_{\rm IF}\propto\Phi^{-1/3}\dot{M}^{2/3}d^{2/3}$, where $\Phi$ is the ionizing flux, $\dot{M}$ is the mass-loss rate, and $d$ is the separation between the proplyd and the ionizing source. The strong ionizing flux from $\theta^1$ Ori C likely sculpted HH 508's small IF.

\subsection*{\textit 3.3. LV 1 \lowercase{in} H$\alpha$}
LV 1 is a binary proplyd separated by $\sim$$0\farcs4$ and exhibits crescent IFs and cometary tails driven by $\theta^1$ Ori C (Figure \ref{fig:HSTvsMagAO}). Similar to $\theta^1$ Ori $\rm{B}_2\rm{B}_3$, both proplyds also have relatively small IFs, with $R_{\rm IF} \sim 40$ au for the major proplyd (168--326 SE) and $\sim$20 au for the minor proplyd (168--326 NW). A strip of diffuse gas between the proplyds has been interpreted as the collision of photoevaporated disk flows \citep{H02}. 

In Figure \ref{fig:LV1} we present the 2012, 2014, and 2016 MagAO H$\alpha$ observations of LV 1. We reveal finer structure with a Gaussian kernel of FWHM = 10 pixels (bottom panels). A bright H$\alpha$ blob at $\sim$0\farcs13 ($\sim$50 au) to the south of the major proplyd appears in the 2014 image, but disappears in our 2016 observation. While its nature remains enigmatic, the blob could be an ejecta from the protostar embedded in the major proplyd. As the blob moves farther away, it disperses and fades away quickly.

\acknowledgements
We are grateful to the reviewer, John Bally, for very helpful comments. We also thank Megan Reiter for discussions. This material is based upon work supported by the National Science Foundation under grant No. 1506818 (PI Males) and NSF AAG grant No. 1615408 (PI Close). Y.-L.W. and L.M.C. are supported by the NSF AAG award and the TRIF fellowship. K.M.M.'s and L.M.C.'s work is supported by the NASA Exoplanets Research Program (XRP) by cooperative agreement NNX16AD44G. This paper includes data gathered with the 6.5 m {\it Magellan} Clay Telescope at Las Campanas Observatory, Chile.


\begin{thebibliography}{}
\bibitem[Allen et al.(2015)]{A15}Allen, C., Costero, R., \& Hern\'{a}ndez, M. 2015, \aj, 150, 167
\bibitem[Allen et al.(2017)]{A17}Allen, C., Costero, R., Ruelas-Mayorga, A., \& S\'{a}nchez, L. J. 2017, \mnras, 466, 4937
\bibitem[Bally et al.(2006)]{B06}Bally, J., Licht, D., Smith, N., \& Walawender, J. 2006, \aj, 131, 473
\bibitem[Bally et al.(2000)]{B00}Bally, J., O'Dell, C. R., \& McCaughrean, M. 2000, \aj, 119, 2919
\bibitem[Bally \& Reipurth(2001)]{BR01}Bally, J., \& Reipurth, B. 2001, \apj, 546, 299
\bibitem[Bally et al.(1998)]{B98}Bally, J., Sutherland, R. S., Devine, D., \& Johnstone, D. 1998, \aj, 116, 293
\bibitem[Castelli \& Kurucz(2004)]{CK04}Castelli, F., \& Kurucz, R. L. 2004, arXiv:astro-ph/0405087
\bibitem[Close et al.(2014)]{C14}Close, L. M., Follette, K. B., Males, J. R., et al. 2014, \apjl, 781, L30
\bibitem[Close et al.(2012)]{C12}Close, L. M., Males, J. R., Kopon, D., et al. 2012, Proc. SPIE, 8447, 84470X
\bibitem[Close et al.(2013)]{C13}Close, L. M., Males, J. R., Morzinski, K., et al. 2013, \apj, 774, 94
\bibitem[Close et al.(2003)]{C03}Close, L. M., Wildi, F., Lloyd-Hart, M., et al. 2003, \apj, 599, 537
\bibitem[Eisner et al.(2016)]{E16}Eisner, J. A., Bally, J. M., Ginsburg, A., \& Sheehan, P. D. 2016, \apj, 826, 16
\bibitem[Getman et al.(2005)]{G05}Getman, K. V., Flaccomio, E., Broos, P. S., et al. 2005, \apjs, 160, 319
\bibitem[Henney(2002)]{H02}Henney, W. J. 2002, RMxAA, 38, 71
\bibitem[Johnstone et al.(1998)]{J98}Johnstone, D., Hollenbach, D., \& Bally, J. 1998, \apj, 499, 758
\bibitem[Kim et al.(2016)]{K16}Kim, J. S., Clarke, C. J., Fang, M., \& Facchini, S. 2016, \apjl, 826, L15
\bibitem[Kounkel et al.(2017)]{K17}Kounkel, M., Hartmann, L., Loinard, L., et al. 2017, \apj, 834, 142
\bibitem[Males et al.(2014)]{Males14}Males, J. R., Close, L. M., Morzinski, K., et al. 2014, \apj, 786, 32
\bibitem[Mann et al.(2014)]{Mann14}Mann, R. K., Di Francesco, J., Johnstone, D., et al. 2014, \apj, 784, 82
\bibitem[Mann \& Williams(2010)]{MW10}Mann, R. K., \& Williams, J. P. 2010, \apj, 725, 430
\bibitem[Morzinski et al.(2014)]{M14}Morzinski, K. M., Close, L. M., Males, J. R., et al. 2014, Proc. SPIE, 9148, 914804
\bibitem[O'Dell \& Wen(1994)]{OW94}O'Dell, C. R., \& Wen, Z. 1994, \apj, 436, 194
\bibitem[O'Dell et al.(1993)]{O93}O'Dell, C. R., Wen, Z., \& Hu, X. 1993, \apj, 410, 696
\bibitem[Petr et al.(1998)]{P98}Petr, M. G., Du Foresto, V., Beckwith, S. V. W., Richichi, A., \& McCaughrean, M. J. 1998, \apj, 500, 825
\bibitem[Reiter et al.(2016)]{R16}Reiter, M., Smith, N., \& Bally, J. 2016, \mnras, 463, 4344
\bibitem[Reiter et al.(2015a)]{R15a}Reiter, M., Smith, N., Kiminki, M. M., \& Bally, J. 2015a, \mnras, 450, 564
\bibitem[Reiter et al.(2015b)]{R15b}Reiter, M., Smith, N., Kiminki, M. M., Bally, J., \& Anderson, J. 2015b, \mnras, 448, 3429
\bibitem[Ricci et al.(2008)]{R08}Ricci, L., Robberto, M., \& Soderblom, D. R. 2008, \aj, 136, 2136
\bibitem[Sheehan et al.(2016)]{S16}Sheehan, P. D., Eisner, J. A., Mann, R. K., \& Williams, J. P. 2016, \apj, 831, 155
\bibitem[Smith et al.(2005)]{Smith05}Smith, N., Bally, J., Shuping, R. Y., Morris, M., \& Kassis, M. 2005, \aj, 130, 1763
\bibitem[Stelzer et al.(2005)]{S05}Stelzer, B., Flaccomio, E., Montmerle, T., et al. 2005, \apjs, 160, 557
\bibitem[Tody(1986)]{T86}Tody, D. 1986, Proc. SPIE, 627, 733
\bibitem[Tody(1993)]{T93}Tody, D. 1993, in ASP Conf. Ser. 52, Astronomical Data Analysis Software and Systems II, ed. R. J. Hanisch, R. J. V. Brissenden, \& J. Barnes (San Francisco, CA: ASP), 173
\bibitem[Vicente \& Alves(2005)]{VA05}Vicente, S. M., \& Alves, J. 2005, \aap, 441, 195
\bibitem[Vitrichenko \& Klochkova(2004)]{VK04}Vitrichenko, \'{E}. A., \& Klochkova, V. G. 2004, Ap, 47, 169 
\bibitem[Vitrichenko et al.(2006)]{V06}Vitrichenko, \'{E}. A., Klochkova, V. G., \& Tsymbal, V. V. 2006, Ap, 49, 96
\bibitem[Windemuth et al.(2013)]{Windemuth13}Windemuth, D., Herbst, W., Tingle, E., et al. 2013, \apj, 768, 67
\bibitem[Wu et al.(2013)]{W13}Wu, Y.-L., Close, L. M., Males, J. R., et al. 2013, \apj, 774, 45
\end{thebibliography}
\end{document}